\documentclass[conference,10pt,letterpaper,nofonttune]{IEEEtran}
\pdfoutput=1
\IEEEoverridecommandlockouts
\usepackage{cite}
\usepackage{csquotes}
\usepackage{amsmath,amssymb,amsfonts}
\usepackage{algorithmic,algorithm}
\usepackage{xcolor}
\usepackage{subfigure}
\usepackage{comment}
\usepackage{xspace}
\usepackage{balance}
\usepackage{fancyhdr}
\usepackage{bm} 

\newcommand{\mycomment}[1]{}
\newcommand{\oran}{O-RAN\xspace}

\usepackage{soul}
\usepackage{multirow}
\usepackage{booktabs}
\usepackage{tabularx}
\usepackage{tikz}
\usepackage{pgfplots}
\usepackage{adjustbox}
\pgfplotsset{compat=newest}
\pgfplotsset{plot coordinates/math parser=false}
\newlength\fheight
\newlength\fwidth
\usetikzlibrary{plotmarks,patterns,decorations.pathreplacing,backgrounds,calc,arrows,arrows.meta,spy,matrix,scopes}
\usepgfplotslibrary{patchplots,groupplots}
\usepackage{tikzscale}
\usepackage[draft]{hyperref}

\newif\ifexttikz
\exttikzfalse

\ifexttikz
	\usetikzlibrary{external}
	\usepackage{fontspec}
\fi

\usepackage[acronyms,nonumberlist,nopostdot,nomain,nogroupskip,acronymlists={hidden}]{glossaries}
\newglossary[algh]{hidden}{acrh}{acnh}{Hidden Acronyms}
\newacronym{3gpp}{3GPP}{3rd Generation Partnership Project}
\newacronym{4g}{4G}{4th generation}
\newacronym{5g}{5G}{5th generation}
\newacronym{6g}{6G}{6th generation}
\newacronym{5gc}{5GC}{5G Core}
\newacronym{adc}{ADC}{Analog to Digital Converter}
\newacronym{aerpaw}{AERPAW}{Aerial Experimentation and Research Platform for Advanced Wireless}
\newacronym{ai}{AI}{Artificial Intelligence}
\newacronym{aimd}{AIMD}{Additive Increase Multiplicative Decrease}
\newacronym{am}{AM}{Acknowledged Mode}
\newacronym{amc}{AMC}{Adaptive Modulation and Coding}
\newacronym{amf}{AMF}{Access and Mobility Management Function}
\newacronym{aops}{AOPS}{Adaptive Order Prediction Scheduling}
\newacronym{api}{API}{Application Programming Interface}
\newacronym{xapp}{xApp}{Intelligent Application}
\newacronym{apn}{APN}{Access Point Name}
\newacronym{aqm}{AQM}{Active Queue Management}
\newacronym{ausf}{AUSF}{Authentication Server Function}
\newacronym{avc}{AVC}{Advanced Video Coding}
\newacronym{awgn}{AGWN}{Additive White Gaussian Noise}
\newacronym{balia}{BALIA}{Balanced Link Adaptation Algorithm}
\newacronym{bbu}{BBU}{Base Band Unit}
\newacronym{bdp}{BDP}{Bandwidth-Delay Product}
\newacronym{ber}{BER}{Bit Error Rate}
\newacronym{bf}{BF}{Beamforming}
\newacronym{bler}{BLER}{Block Error Rate}
\newacronym{brr}{BRR}{Bayesian Ridge Regressor}
\newacronym{bs}{BS}{Base Station}
\newacronym{bsr}{BSR}{Buffer Status Report}
\newacronym{bss}{BSS}{Business Support System}
\newacronym{ca}{CA}{Carrier Aggregation}
\newacronym{caas}{CaaS}{Connectivity-as-a-Service}
\newacronym{cb}{CB}{Code Block}
\newacronym{cc}{CC}{Congestion Control}
\newacronym{ccid}{CCID}{Congestion Control ID}
\newacronym{cv2x}{C-V2X}{Cellular Vehicle-to-Everything}
\newacronym{cco}{CC}{Carrier Component}
\newacronym{cdd}{CDD}{Cyclic Delay Diversity}
\newacronym{cdf}{CDF}{Cumulative Distribution Function}
\newacronym{cdn}{CDN}{Content Distribution Network}
\newacronym{cn}{CN}{Core Network}
\newacronym{codel}{CoDel}{Controlled Delay Management}
\newacronym{comac}{COMAC}{Converged Multi-Access and Core}
\newacronym{cord}{CORD}{Central Office Re-architected as a Datacenter}
\newacronym{cornet}{CORNET}{COgnitive Radio NETwork}
\newacronym{cosmos}{COSMOS}{Cloud Enhanced Open Software Defined Mobile Wireless Testbed for City-Scale Deployment}
\newacronym{cots}{COTS}{Commercial Off-the-Shelf}
\newacronym{cp}{CP}{Control Plane}
\newacronym{cpu}{CPU}{Central Processing Unit}
\newacronym{cqi}{CQI}{Channel Quality Information}
\newacronym{cr}{CR}{Cognitive Radio}
\newacronym{cran}{CRAN}{Cloud \gls{ran}}
\newacronym{crs}{CRS}{Cell Reference Signal}
\newacronym{csi}{CSI}{Channel State Information}
\newacronym{csirs}{CSI-RS}{Channel State Information - Reference Signal}
\newacronym{cu}{CU}{Central Unit}
\newacronym{d2tcp}{D$^2$TCP}{Deadline-aware Data center TCP}
\newacronym{d3}{D$^3$}{Deadline-Driven Delivery}
\newacronym{dac}{DAC}{Digital to Analog Converter}
\newacronym{dag}{DAG}{Directed Acyclic Graph}
\newacronym{das}{DAS}{Distributed Antenna System}
\newacronym{dash}{DASH}{Dynamic Adaptive Streaming over HTTP}
\newacronym{dc}{DC}{Dual Connectivity}
\newacronym{dccp}{DCCP}{Datagram Congestion Control Protocol}
\newacronym{dce}{DCE}{Direct Code Execution}
\newacronym{dci}{DCI}{Downlink Control Information}
\newacronym{dctcp}{DCTCP}{Data Center TCP}
\newacronym{dl}{DL}{Downlink}
\newacronym{dmr}{DMR}{Deadline Miss Ratio}
\newacronym{dmrs}{DMRS}{DeModulation Reference Signal}
\newacronym{drlcc}{DRL-CC}{Deep Reinforcement Learning Congestion Control}
\newacronym{drs}{DRS}{Discovery Reference Signal}
\newacronym{du}{DU}{Distributed Unit}
\newacronym{e2e}{E2E}{end-to-end}
\newacronym{ecaas}{ECaaS}{Edge-Cloud-as-a-Service}
\newacronym{ecn}{ECN}{Explicit Congestion Notification}
\newacronym{edf}{EDF}{Earliest Deadline First}
\newacronym{embb}{eMBB}{Enhanced Mobile Broadband}
\newacronym{empower}{EMPOWER}{EMpowering transatlantic PlatfOrms for advanced WirEless Research}
\newacronym{enb}{eNB}{evolved Node Base}
\newacronym{endc}{EN-DC}{E-UTRAN-\gls{nr} \gls{dc}}
\newacronym{epc}{EPC}{Evolved Packet Core}
\newacronym{eps}{EPS}{Evolved Packet System}
\newacronym{es}{ES}{Edge Server}
\newacronym{etsi}{ETSI}{European Telecommunications Standards Institute}
\newacronym[firstplural=Estimated Times of Arrival (ETAs)]{eta}{ETA}{Estimated Time of Arrival}
\newacronym{eutran}{E-UTRAN}{Evolved Universal Terrestrial Access Network}
\newacronym{faas}{FaaS}{Function-as-a-Service}
\newacronym{fapi}{FAPI}{Functional Application Platform Interface}
\newacronym{fdd}{FDD}{Frequency Division Duplexing}
\newacronym{fdm}{FDM}{Frequency Division Multiplexing}
\newacronym{fdma}{FDMA}{Frequency Division Multiple Access}
\newacronym{fed4fire}{FED4FIRE+}{Federation 4 Future Internet Research and Experimentation Plus}
\newacronym{fir}{FIR}{Finite Impulse Response}
\newacronym{cir}{CIR}{Channel Impulse Response}
\newacronym{fit}{FIT}{Future \acrlong{iot}}
\newacronym{fpga}{FPGA}{Field Programmable Gate Array}
\newacronym{fr2}{FR2}{Frequency Range 2}
\newacronym{fr1}{FR1}{Frequency Range 1}
\newacronym{fs}{FS}{Fast Switching}
\newacronym{fscc}{FSCC}{Flow Sharing Congestion Control}
\newacronym{ftp}{FTP}{File Transfer Protocol}
\newacronym{fw}{FW}{Flow Window}
\newacronym{ge}{GE}{Gaussian Elimination}
\newacronym{gnb}{gNB}{Next Generation Node Base}
\newacronym{nextg}{NextG}{Next Generation}
\newacronym{gop}{GOP}{Group of Pictures}
\newacronym{gpr}{GPR}{Gaussian Process Regressor}
\newacronym{gpu}{GPU}{Graphics Processing Unit}
\newacronym{gtp}{GTP}{GPRS Tunneling Protocol}
\newacronym{gtpc}{GTP-C}{GPRS Tunnelling Protocol Control Plane}
\newacronym{sca}{SCA}{Successive Convex Approximation}
\newacronym{gtpu}{GTP-U}{GPRS Tunnelling Protocol User Plane}
\newacronym{gtpv2c}{GTPv2-C}{\gls{gtp} v2 - Control}
\newacronym{gw}{GW}{Gateway}
\newacronym{harq}{HARQ}{Hybrid Automatic Repeat reQuest}
\newacronym{hetnet}{HetNet}{Heterogeneous Network}
\newacronym{hh}{HH}{Hard Handover}
\newacronym{hol}{HOL}{Head-of-Line}
\newacronym{hqf}{HQF}{Highest-quality-first}
\newacronym{hss}{HSS}{Home Subscription Server}
\newacronym{http}{HTTP}{HyperText Transfer Protocol}
\newacronym{ia}{IA}{Initial Access}
\newacronym{iab}{IAB}{Integrated Access and Backhaul}
\newacronym{ic}{IC}{Incident Command}
\newacronym{ietf}{IETF}{Internet Engineering Task Force}
\newacronym{imsi}{IMSI}{International Mobile Subscriber Identity}
\newacronym{imt}{IMT}{International Mobile Telecommunication}
\newacronym{iot}{IoT}{Internet of Things}
\newacronym{ip}{IP}{Internet Protocol}
\newacronym{itu}{ITU}{International Telecommunication Union}
\newacronym{kpi}{KPI}{Key Performance Indicator}
\newacronym{kpm}{KPM}{Key Performance Measurement}
\newacronym{kvm}{KVM}{Kernel-based Virtual Machine}
\newacronym{los}{LOS}{Line-of-Sight}
\newacronym{lsm}{LSM}{Link-to-System Mapping}
\newacronym{lstm}{LSTM}{Long Short Term Memory}
\newacronym{lte}{LTE}{Long Term Evolution}
\newacronym{lxc}{LXC}{Linux Container}
\newacronym{m2m}{M2M}{Machine to Machine}
\newacronym{mac}{MAC}{Medium Access Control}
\newacronym{manet}{MANET}{Mobile Ad Hoc Network}
\newacronym{mano}{MANO}{Management and Orchestration}
\newacronym{mc}{MC}{Multi-Connectivity}
\newacronym{mcc}{MCC}{Mobile Cloud Computing}
\newacronym{mchem}{MCHEM}{Massive Channel Emulator}
\newacronym{mcs}{MCS}{Modulation and Coding Scheme}
\newacronym{mec}{MEC}{Multi-access Edge Computing}
\newacronym{mec2}{MEC}{Mobile Edge Cloud}
\newacronym{mfc}{MFC}{Mobile Fog Computing}
\newacronym{mgen}{MGEN}{Multi-Generator}
\newacronym{mi}{MI}{Mutual Information}
\newacronym{mib}{MIB}{Master Information Block}
\newacronym{miesm}{MIESM}{Mutual Information Based Effective SINR}
\newacronym{mimo}{MIMO}{Multiple Input, Multiple Output}
\newacronym{ml}{ML}{Machine Learning}
\newacronym{mlr}{MLR}{Maximum-local-rate}
\newacronym[plural=\gls{mme}s,firstplural=Mobility Management Entities (MMEs)]{mme}{MME}{Mobility Management Entity}
\newacronym{mmtc}{mMTC}{Massive Machine-Type Communications}
\newacronym{mmwave}{mmWave}{millimeter wave}
\newacronym{mpdccp}{MP-DCCP}{Multipath Datagram Congestion Control Protocol}
\newacronym{mptcp}{MPTCP}{Multipath TCP}
\newacronym{mr}{MR}{Maximum Rate}
\newacronym{mrdc}{MR-DC}{Multi \gls{rat} \gls{dc}}
\newacronym{mse}{MSE}{Mean Square Error}
\newacronym{mss}{MSS}{Maximum Segment Size}
\newacronym{mt}{MT}{Mobile Termination}
\newacronym{mtd}{MTD}{Machine-Type Device}
\newacronym{mtu}{MTU}{Maximum Transmission Unit}
\newacronym{mumimo}{MU-MIMO}{Multi-user \gls{mimo}}
\newacronym{mvno}{MVNO}{Mobile Virtual Network Operator}
\newacronym{nalu}{NALU}{Network Abstraction Layer Unit}
\newacronym{nas}{NAS}{Non-Access Stratum}
\newacronym{nbiot}{NB-IoT}{Narrow Band IoT}
\newacronym{nfv}{NFV}{Network Function Virtualization}
\newacronym{nfvi}{NFVI}{Network Function Virtualization Infrastructure}
\newacronym{nic}{NIC}{Network Interface Card}
\newacronym{nlos}{NLOS}{Non-Line-of-Sight}
\newacronym{now}{NOW}{Non Overlapping Window}
\newacronym{nsm}{NSM}{Network Service Mesh}
\newacronym[type=hidden]{nr}{NR}{New Radio}
\newacronym{nrf}{NRF}{Network Repository Function}
\newacronym{nsa}{NSA}{Non Stand Alone}
\newacronym{nse}{NSE}{Network Slicing Engine}
\newacronym{nssf}{NSSF}{Network Slice Selection Function}
\newacronym{o2i}{O2I}{Outdoor to Indoor}
\newacronym{oai}{OAI}{OpenAirInterface}
\newacronym{oaicn}{OAI-CN}{\gls{oai} \acrlong{cn}}
\newacronym{oairan}{OAI-RAN}{\acrlong{oai} \acrlong{ran}}
\newacronym{oam}{OAM}{Operations, Administration and Maintenance}
\newacronym{ofdm}{OFDM}{Orthogonal Frequency Division Multiplexing}
\newacronym{olia}{OLIA}{Opportunistic Linked Increase Algorithm}
\newacronym{omec}{OMEC}{Open Mobile Evolved Core}
\newacronym{onap}{ONAP}{Open Network Automation Platform}
\newacronym{onf}{ONF}{Open Networking Foundation}
\newacronym{onos}{ONOS}{Open Networking Operating System}
\newacronym{oom}{OOM}{\gls{onap} Operations Manager}
\newacronym{opnfv}{OPNFV}{Open Platform for \gls{nfv}}
\newacronym{orbit}{ORBIT}{Open-Access Research Testbed for Next-Generation Wireless Networks}
\newacronym{os}{OS}{Operating System}
\newacronym{oss}{OSS}{Operations Support System}
\newacronym{pa}{PA}{Position-aware}
\newacronym{pase}{PASE}{Prioritization, Arbitration, and Self-adjusting Endpoints}
\newacronym{pawr}{PAWR}{Platforms for Advanced Wireless Research}
\newacronym{pbch}{PBCH}{Physical Broadcast Channel}
\newacronym{pcef}{PCEF}{Policy and Charging Enforcement Function}
\newacronym{pcfich}{PCFICH}{Physical Control Format Indicator Channel}
\newacronym{pcrf}{PCRF}{Policy and Charging Rules Function}
\newacronym{pdcch}{PDCCH}{Physical Downlink Control Channel}
\newacronym{pdcp}{PDCP}{Packet Data Convergence Protocol}
\newacronym{pdsch}{PDSCH}{Physical Downlink Shared Channel}
\newacronym{pdu}{PDU}{Packet Data Unit}
\newacronym{pf}{PF}{Proportionally Fair}
\newacronym{pgw}{PGW}{Packet Gateway}
\newacronym{phich}{PHICH}{Physical Hybrid ARQ Indicator Channel}
\newacronym{phy}{PHY}{Physical}
\newacronym{pmch}{PMCH}{Physical Multicast Channel}
\newacronym{pmi}{PMI}{Precoding Matrix Indicators}
\newacronym{powder}{POWDER}{Platform for Open Wireless Data-driven Experimental Research}
\newacronym{ppo}{PPO}{Proximal Policy Optimization}
\newacronym{ppp}{PPP}{Poisson Point Process}
\newacronym{prach}{PRACH}{Physical Random Access Channel}
\newacronym{prb}{PRB}{Physical Resource Block}
\newacronym{rbg}{RBG}{Resource Block Group}
\newacronym{psnr}{PSNR}{Peak Signal to Noise Ratio}
\newacronym{pss}{PSS}{Primary Synchronization Signal}
\newacronym{pucch}{PUCCH}{Physical Uplink Control Channel}
\newacronym{pusch}{PUSCH}{Physical Uplink Shared Channel}
\newacronym{qam}{QAM}{Quadrature Amplitude Modulation}
\newacronym{qci}{QCI}{\gls{qos} Class Identifier}
\newacronym{qoe}{QoE}{Quality of Experience}
\newacronym{qos}{QoS}{Quality of Service}
\newacronym{quic}{QUIC}{Quick UDP Internet Connections}
\newacronym{rach}{RACH}{Random Access Channel}
\newacronym{ran}{RAN}{Radio Access Network}
\newacronym[firstplural=end to endcess Technologies (RATs)]{rat}{RAT}{end to endcess Technology}
\newacronym{rcn}{RCN}{Research Coordination Network}
\newacronym{rec}{REC}{Radio Edge Cloud}
\newacronym{ra}{RA}{Resource Allocation}
\newacronym{red}{RED}{Random Early Detection}
\newacronym{renew}{RENEW}{Reconfigurable Eco-system for Next-generation End-to-end Wireless}
\newacronym{rf}{RF}{Radio Frequency}
\newacronym{rfc}{RFC}{Request for Comments}
\newacronym{rfr}{RFR}{Random Forest Regressor}
\newacronym{ric}{RIC}{\gls{ran} Intelligent Controller}
\newacronym{rlc}{RLC}{Radio Link Control}
\newacronym{rlf}{RLF}{Radio Link Failure}
\newacronym{rlnc}{RLNC}{Random Linear Network Coding}
\newacronym{rmr}{RMR}{RIC Message Router}
\newacronym{rmse}{RMSE}{Root Mean Squared Error}
\newacronym{rnis}{RNIS}{Radio Network Information Service}
\newacronym{rr}{RR}{Round Robin}
\newacronym{rrc}{RRC}{Radio Resource Control}
\newacronym{rrm}{RRM}{Radio Resource Management}
\newacronym{rru}{RRU}{Remote Radio Unit}
\newacronym{rs}{RS}{Remote Server}
\newacronym{rsrp}{RSRP}{Reference Signal Received Power}
\newacronym{rsrq}{RSRQ}{Reference Signal Received Quality}
\newacronym{rss}{RSS}{Received Signal Strength}
\newacronym{rssi}{RSSI}{Received Signal Strength Indicator}
\newacronym{rtt}{RTT}{Round Trip Time}
\newacronym{ru}{RU}{Radio Unit}
\newacronym{rw}{RW}{Receive Window}
\newacronym{rx}{RX}{Receiver}
\newacronym{s1ap}{S1AP}{S1 Application Protocol}
\newacronym{sa}{SA}{standalone}
\newacronym{sack}{SACK}{Selective Acknowledgment}
\newacronym{sap}{SAP}{Service Access Point}
\newacronym{sc2}{SC2}{Spectrum Collaboration Challenge}
\newacronym{scef}{SCEF}{Service Capability Exposure Function}
\newacronym{sch}{SCH}{Secondary Cell Handover}
\newacronym{scoot}{SCOOT}{Split Cycle Offset Optimization Technique}
\newacronym{sctp}{SCTP}{Stream Control Transmission Protocol}
\newacronym{sdap}{SDAP}{Service Data Adaptation Protocol}
\newacronym{sdk}{SDK}{Software Development Kit}
\newacronym{sdm}{SDM}{Space Division Multiplexing}
\newacronym{sdma}{SDMA}{Spatial Division Multiple Access}
\newacronym{sdn}{SDN}{Software-defined Networking}
\newacronym{sdr}{SDR}{Software-defined Radio}
\newacronym{seba}{SEBA}{SDN-Enabled Broadband Access}
\newacronym{sgsn}{SGSN}{Serving GPRS Support Node}
\newacronym{sgw}{SGW}{Service Gateway}
\newacronym{si}{SI}{Study Item}
\newacronym{sib}{SIB}{Secondary Information Block}
\newacronym{sinr}{SINR}{Signal to Interference plus Noise Ratio}
\newacronym{sip}{SIP}{Session Initiation Protocol}
\newacronym{siso}{SISO}{Single Input, Single Output}
\newacronym{sla}{SLA}{Service Level Agreement}
\newacronym{sm}{SM}{Service Model}
\newacronym{smf}{SMF}{Session Management Function}
\newacronym{smo}{SMO}{Service Management and Orchestration}
\newacronym{sms}{SMS}{Short Message Service}
\newacronym{smsgmsc}{SMS-GMSC}{\gls{sms}-Gateway}
\newacronym{snr}{SNR}{Signal-to-Noise-Ratio}
\newacronym{son}{SON}{Self-Organizing Network}
\newacronym{sptcp}{SPTCP}{Single Path TCP}
\newacronym{srb}{SRB}{Service Radio Bearer}
\newacronym{srn}{SRN}{Standard Radio Node}
\newacronym{srs}{SRS}{Sounding Reference Signal}
\newacronym{ss}{SS}{Synchronization Signal}
\newacronym{sss}{SSS}{Secondary Synchronization Signal}
\newacronym{st}{ST}{Spanning Tree}
\newacronym{svc}{SVC}{Scalable Video Coding}
\newacronym{tb}{TB}{Transport Block}
\newacronym{tcp}{TCP}{Transmission Control Protocol}
\newacronym{tdd}{TDD}{Time Division Duplexing}
\newacronym{tdm}{TDM}{Time Division Multiplexing}
\newacronym{tdma}{TDMA}{Time Division Multiple Access}
\newacronym{tfl}{TfL}{Transport for London}
\newacronym{tfrc}{TFRC}{TCP-Friendly Rate Control}
\newacronym{tft}{TFT}{Traffic Flow Template}
\newacronym{tgen}{TGEN}{Traffic Generator}
\newacronym{tip}{TIP}{Telecom Infra Project}
\newacronym{tm}{TM}{Transparent Mode}
\newacronym{to}{TO}{Telco Operator}
\newacronym{tr}{TR}{Technical Report}
\newacronym{trp}{TRP}{Transmitter Receiver Pair}
\newacronym{ts}{TS}{Technical Specification}
\newacronym{tti}{TTI}{Transmission Time Interval}
\newacronym{ttt}{TTT}{Time-to-Trigger}
\newacronym{tx}{TX}{Transmitter}
\newacronym{uas}{UAS}{Unmanned Aerial System}
\newacronym{uav}{UAV}{Unmanned Aerial Vehicle}
\newacronym{udm}{UDM}{Unified Data Management}
\newacronym{udp}{UDP}{User Datagram Protocol}
\newacronym{udr}{UDR}{Unified Data Repository}
\newacronym{ue}{UE}{User Equipment}
\newacronym{uhd}{UHD}{\gls{usrp} Hardware Driver}
\newacronym{ul}{UL}{Uplink}
\newacronym{ap}{AP}{Access Point}
\newacronym{um}{UM}{Unacknowledged Mode}
\newacronym{uml}{UML}{Unified Modeling Language}
\newacronym{upa}{UPA}{Uniform Planar Array}
\newacronym{upf}{UPF}{User Plane Function}
\newacronym{urllc}{URLLC}{Ultra Reliable and Low Latency Communications}
\newacronym{usa}{U.S.}{United States}
\newacronym{usim}{USIM}{Universal Subscriber Identity Module}
\newacronym{usrp}{USRP}{Universal Software Radio Peripheral}
\newacronym{utc}{UTC}{Urban Traffic Control}
\newacronym{vim}{VIM}{Virtualization Infrastructure Manager}
\newacronym{vm}{VM}{Virtual Machine}
\newacronym{vnf}{VNF}{Virtual Network Function}
\newacronym{volte}{VoLTE}{Voice over \gls{lte}}
\newacronym{voltha}{VOLTHA}{Virtual OLT HArdware Abstraction}
\newacronym{vr}{VR}{Virtual Reality}
\newacronym{vran}{vRAN}{Virtualized \gls{ran}}
\newacronym{vss}{VSS}{Video Streaming Server}
\newacronym{wbf}{WBF}{Wired Bias Function}
\newacronym{wf}{WF}{Waterfilling}
\newacronym{wlan}{WLAN}{Wireless Local Area Network}
\newacronym{osm}{OSM}{Open Source \gls{nfv} Management and Orchestration}
\newacronym{pnf}{PNF}{Physical Network Function}
\newacronym{drl}{DRL}{Deep Reinforcement Learning}
\newacronym{rl}{RL}{Reinforcement Learning}
\newacronym{mtc}{MTC}{Machine-type Communications}
\newacronym{osc}{OSC}{O-RAN Software Community}
\newacronym{rc}{RC}{RAN Control}
\newacronym{dqn}{DQN}{Deep Q-Network}
\newacronym{v2x}{V2X}{Vehicle-to-everything}
\newacronym{gbsm}{GBSM}{Geometry-Based Stochastic Model}
\newacronym{gbs}{GBSM}{Geometry-Based Stochastic}
\newacronym{quadriga}{QuaDRiGa}{QUAsi Deterministic RadIo channel GenerAtor}
\newacronym{relu}{ReLU}{Rectified Linear Unit} 
\newacronym{mpc}{MPC}{Multipath Component}
\newacronym{xpr}{XPR}{Cross-polarization Ratio}
\newacronym{lsp}{LSP}{Large Scale Parameter}
\newacronym{ssp}{SSP}{Small Scale Parameter}
\newacronym{fbs}{FBS}{First Bounce Scatterer}
\newacronym{lbs}{LBS}{Last Bounce Scatterer}
\newacronym{d2d}{D2D}{Device-to-Device}
\newacronym{rsu}{RSU}{Road Side Unit}
\newacronym{toa}{ToA}{Time-of-Arrival}
\newacronym{ris}{RIS}{Reconfigurable Intelligent Surface}
\newacronym{aoa}{AoA}{Angle of Arrival}
\newacronym{aod}{AoD}{Angle of Departure}
\newacronym{pl}{PL}{Path-Loss}
\newacronym{noma}{NOMA}{Non-Orthogonal Multiple Access}
\newacronym{sic}{SIC}{Successive Interference Cancellation}
\newacronym{gps}{GPS}{Global Positioning System}
\newacronym{ids}{IDS}{Independent Diffusive Scatterer-based}
\newacronym{inw}{INW}{Impedance Network-based}
\newacronym{minlp}{MINLP}{Mixed Integer Non-Linear Programming}
\newacronym{star}{STAR}{Simultaneous Transmitting And Reflecting}
\glsdisablehyper

\usepackage{tikzpagenodes,etoolbox}
\usetikzlibrary{calc}
\usepackage[contents={}]{background}
\AddEverypageHook{%
\ifnumequal{\thepage}{1}{%
 \tikz[remember picture,overlay]{%
     \node[draw,
     text width=0.95\textwidth,
     font=\footnotesize
     ]
     at ($(current page header area) - (0,5pt)$)
     {%
    This paper has been accepted for publication on IEEE Workshop on Next-generation Open and Programmable Radio Access Networks (NG-OPERA), 2025. This is the authors’
accepted version of the article. The final version published by IEEE is: M. Tsampazi, M. Polese, F. Dressler, T. Melodia, “O-RIS-ing: Evaluating RIS-Assisted
NextG Open RAN,” Proc. of the 3nd IEEE
Workshop on Next-generation Open and Programmable Radio Access Networks (NG-OPERA), London, United Kingdom, May 2025.
     };
 }%
}{}
}

\begin{document}

\title{O-RIS-ing: Evaluating RIS-Assisted \\ NextG Open RAN\vspace{-.3cm}}

\author{\IEEEauthorblockN{Maria Tsampazi\IEEEauthorrefmark{1}, Michele Polese\IEEEauthorrefmark{1}, Falko Dressler\IEEEauthorrefmark{4}, Tommaso Melodia\IEEEauthorrefmark{1}}
\IEEEauthorblockA{\IEEEauthorrefmark{1}Institute for the Wireless Internet of Things, Northeastern University, Boston, MA, U.S.A.\\E-mail: \{tsampazi.m, m.polese, t.melodia\}@northeastern.edu\\\IEEEauthorrefmark{4}School of Electrical Engineering and Computer Science, TU Berlin, Germany\\E-mail: \{dressler\}@ccs-labs.org}
\thanks{This article is based upon work partially supported by the U.S.\ National Science Foundation under grants CNS-1925601 and CNS-2112471.}
}

\makeatletter
\patchcmd{\@maketitle}
  {\addvspace{0.5\baselineskip}\egroup} 
  {\addvspace{-1.75\baselineskip}\egroup} 
  {}
  {}
\makeatother

\maketitle

\glsunset{usrp}

\begin{abstract}
\glspl{ris} pose as a transformative technology to revolutionize the cellular architecture of \gls{nextg} \glspl{ran}. Previous studies have demonstrated the capabilities of \glspl{ris} in optimizing wireless propagation, achieving high spectral efficiency, and improving resource utilization. At the same time, the transition to softwarized, disaggregated, and virtualized architectures, such as those being standardized by the \oran ALLIANCE, enables the vision of a reconfigurable Open \gls{ran}.
In this work, we aim to integrate these technologies by studying how different resource allocation policies enhance the performance of \gls{ris}-assisted Open \glspl{ran}.
We perform a comparative analysis among various network configurations and show how proper network optimization can enhance the performance across the \gls{embb} and \gls{urllc} network slices, achieving up to $\sim34\%$ throughput improvement. Furthermore, leveraging the capabilities of OpenRAN Gym, we deploy an xApp on Colosseum, the world's largest wireless system emulator with hardware-in-the-loop, to control the \gls{bs}'s scheduling policy. Experimental results demonstrate that \gls{ris}-assisted topologies achieve high resource efficiency and low latency, regardless of the \gls{bs}'s scheduling policy.
\end{abstract}

\glsresetall
\glsunset{nextg}
\glsunset{ris}
\glsunset{ran}
\glsunset{enb}
\glsunset{embb}
\glsunset{urllc}
\glsunset{ric}


\section{Introduction}\label{intro}
The Open \gls{ran} paradigm for \oran networks is pivotal in transforming cellular network design. Grounded in the principles of disaggregation, virtualization, and programmable radios, components are interconnected through open, standardized interfaces, ensuring multi-vendor interoperability. This approach enables the deployment of flexible architectures built on cloud-native principles. 
The 
reconfigurability of the \gls{ran} is further enhanced through \glspl{ric}, streaming \gls{ran} telemetry and implementing control policies via a closed-loop mechanism~\cite{polese2023understanding}. Both near- and non-real time \glspl{ric}, deployed at the \gls{ran}'s edge, leverage xApps and rApps to optimize network management and resource allocation. This functionality further advances the integration of \gls{ai} for the \gls{ran} and \gls{ai} on the \gls{ran}, fostering the development of intelligent and adaptive \gls{ran} architectures~\cite{AIRANAlliance2024}.

At the same time, \glspl{ris} are envisioned to be an integral part of \gls{nextg} cellular architectures, enabling smart propagation environments that can be controlled through 
programmable \gls{rf} components~\cite{liaskos2022software}. Thanks to their reconfigurability, \glspl{ris} can manipulate electromagnetic waves, reshaping wireless propagation to optimize spectrum utilization and efficiency, particularly in 
multi-operator networks~\cite{angjo2023side}. This reconfiguration capability aligns with \oran's objectives of developing adaptable and resource-efficient \glspl{ran}. 

Intelligent control loops with xApps in \oran have attracted widespread attention from the research community~\cite{bonati2023openran}. In~\cite{tsampazi2024pandora}, the authors explore the impact of xApps embedding \gls{drl} agents to manage the \gls{bs}'s slicing and scheduling control policies across various network deployments on the Colosseum testbed. In addition, several efforts have been conducted in the optimization of wireless propagation with \glspl{ris}. Some previous works~\cite{li2020reconfigurable,diamanti2021energy,diamanti2021prospect} have focused on the use of \glspl{ris} and constructive beams to achieve coherent signal construction at the \gls{bs}, for maximizing the achieved \gls{rssi}.  Finally, a system-level experimental evaluation of \gls{ris}-enabled deployments and their cross-layer optimization, leveraging the full protocol stack on the Colosseum testbed has been discussed in~\cite{tsampazi2024system}.

\subsection{Contributions and Outline} \label{Section IB}
However, limited research has focused on bridging 
\oran and \glspl{ris} at a system level.
In this study, we aim to address this research gap by evaluating network performance across various \gls{ris}-assisted topologies using the Colosseum Open \gls{ran} testbed~\cite{polese2024colosseum}. We consider an Open \gls{ran} delivering services to the \gls{embb} and \gls{urllc} network slices. Using OpenRAN Gym~\cite{bonati2023openran}---an open-source framework for experimentation in \oran---we deploy this Open \gls{ran} on the Colosseum network emulator and manage it through an xApp that controls the \gls{bs}'s scheduling policy. By investigating various spectrum allocation policies and \gls{ris} deployments, we demonstrate how network performance and resource efficiency are impacted differently for each slice. 
With this work, we aim to complement prior studies~\cite{tsampazi2024system} by providing a comprehensive evaluation of \gls{ris}-assisted Open \glspl{ran} on an \oran-compliant testbed.
We believe that our findings and insights will contribute to a deeper understanding of the 
integration of \glspl{ris} in \gls{nextg} Open \glspl{ran}.

The remainder of this paper is organized as follows. Section~\ref{system model} describes the system model. Section~\ref{opt} presents the optimization framework for shaping wireless propagation with \glspl{ris}, while Section~\ref{strategies} discusses the resource allocation strategies. Section~\ref{expsetup} details the experimental setup, while Section~\ref{expevaluation} discusses the experimental results. Finally, Section~\ref{conculsion} draws our conclusions and outlines directions for future work.

\noindent

\section{System Model}\label{system model}

\subsection{Open RAN Framework}\label{oranframe}
\begin{figure}[t!]
  \centering
  \includegraphics[width=\columnwidth]{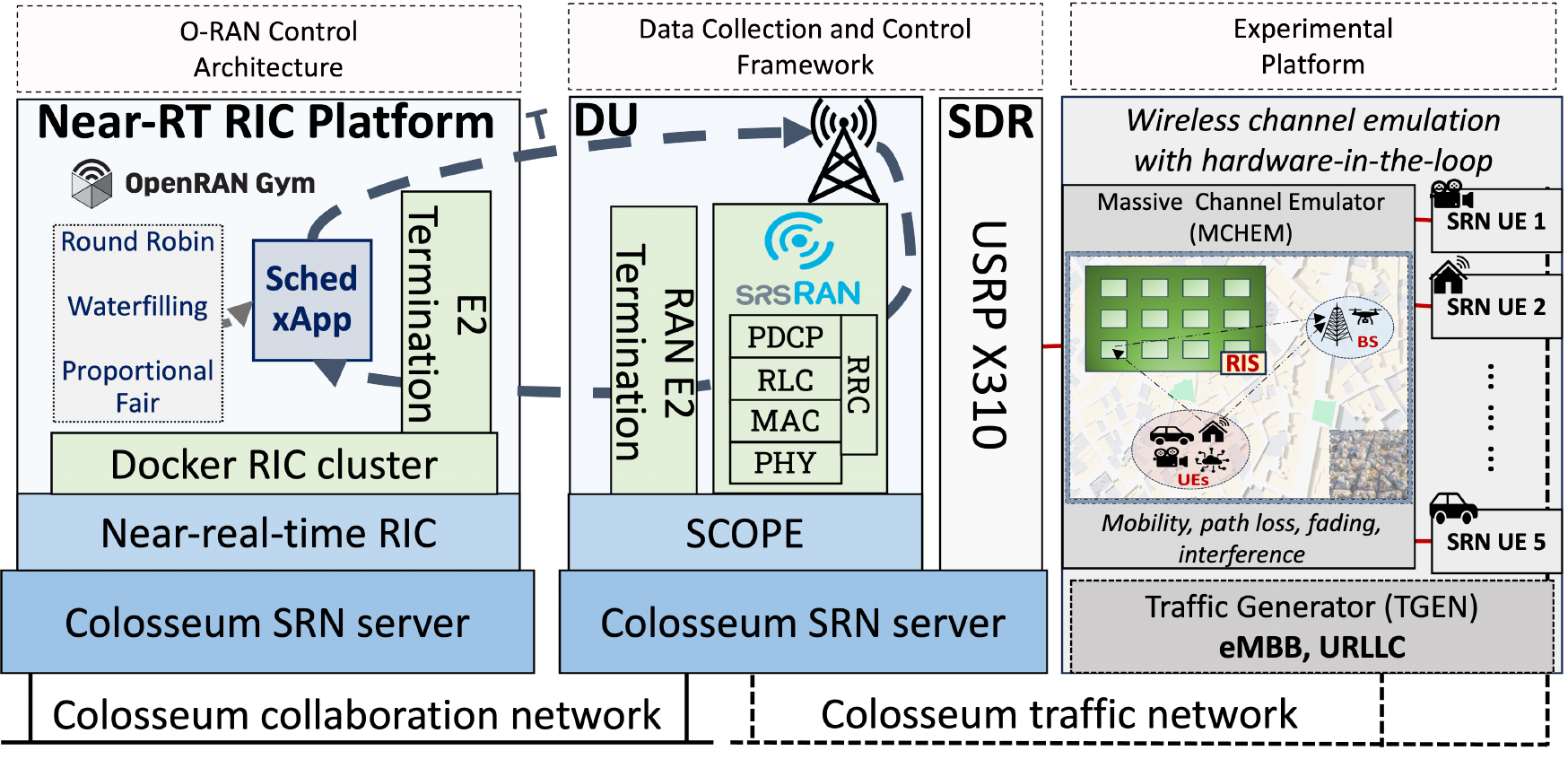} 
  \setlength\abovecaptionskip{-.2cm}
  \caption{Reference \oran testing architecture for \gls{ris}-enabled deployments, focusing on the \texttt{Sched} xApp Case Study, as described in Section~\ref{multiue-secc}.}
  \label{fig:ris-top}
  \vspace{-.9em}
\end{figure}

We investigate an Open \gls{ran} system where \glspl{ue} generate traffic with diverse profiles, which is classified into the \gls{embb} and \gls{urllc} network slices. We leverage an xApp tasked with reconfiguring the \gls{bs}'s \gls{mac}-layer scheduling policies. The xApp selects a dedicated scheduling profile among multiple resource allocation algorithms (i.e., \gls{rr}, \gls{wf}, and \gls{pf}), influencing how \glspl{prb} are internally allocated to the \glspl{ue} of each slice~\cite{tsampazi2024pandora}. It is noted that for the \gls{rf} environment we consider various \gls{ris}-assisted network deployments. 
In Fig.~\ref{fig:ris-top}, we outline the reference architecture considered in our work.

\subsection{Channel Model}\label{channelmod}
We consider a topology consisting of a \gls{ris}, a \gls{bs}, and a set of \glspl{ue}, denoted by $I=\{1,\dots,i,\dots, |I|\}$, which establish direct communication with the \gls{bs}. The number of elements on the \gls{ris} is given as $|M|$ and their set is defined as $M=\{1, \dots , m, \dots , |M|\}$. The first \gls{ris} element, i.e., $m=1$, with its respective coordinates, i.e., $(x_R, y_R, z_R)$ [m], is used as a reference point in the following arithmetic calculations, while for the \gls{bs}, its coordinates are given as follows $(x_A, y_A, z_A)$ [m]. 
The channel gain corresponding to the direct link between a \gls{ue} $i$ and the \gls{bs} is given as $h_{iA}$, the channel gain between a \gls{ue} $i$ and the \gls{ris} is defined as $\mathbf{h}_{iR}$, where $\mathbf{h}_{iR}=[|h_{iR,1}| e^{j \omega_{1}}, \dots,|h_{iR,m}| e^{j \omega_{m}}, \dots,|h_{iR,|M|}| e^{j\omega_{|M|}}]^T$, and finally, the channel gain of the reflected link from the \gls{ris} to the \gls{bs} is denoted as $\mathbf{h}_{RA}$. 
The diagonal phase-shift matrix is defined as $\boldsymbol{\Theta}=\mathrm{diag}( e^{j\theta_1},\dots,e^{j\theta_m},\dots,e^{j\theta_{|M|}} )$, while
each \gls{ris} element's phase shift is $\theta_m \in [0,2\pi]$, $\forall~m~\in~M$, and the cascaded channel is therefore given as $\mathbf{{h}}_{RA}^H\boldsymbol{\Theta}\mathbf{{h}}_{iR}$. Finally, the steering vector for the \gls{ue} $i$ to the \gls{ris} link is given as $\mathbf{h}_{iR}^{\prime\prime\prime}=[1, e^{-j \frac{2 \pi}{\lambda} d \phi_{iR}}, \dots, e^{-j \frac{2 \pi}{\lambda}(|M|-1) d \phi_{i R}}]^{T}$, while the steering vector for the \gls{ris} to the \gls{bs} link is denoted as $\mathbf{h}_{RA}^{\prime\prime\prime}=[1, e^{-j \frac{2 \pi}{\lambda} d \phi_{RA}}, \dots, e^{-j \frac{2 \pi}{\lambda}(|M|-1) d \phi_{RA}}]^T$. In the above calculations, $d$ is the Euclidean distance, $\lambda [m]$ is the carrier wavelength and $d [m]$ is the antenna separation. Additionally, $\phi_{RA}$ is the cosine of the signal's \gls{aod} for the \gls{ris} to the \gls{bs} link, and
$\phi_{iR}$ is the Cosine of the signal's \gls{aoa} for the \gls{ue} $i$ to \gls{ris} link. 
It is noted that we consider the links between the \glspl{ue} and the \gls{bs} to be \gls{nlos}, while the channels connecting the \glspl{ue} to the \gls{ris} and the \gls{ris} to the \gls{bs} are assumed to be \gls{los}. These transmission conditions are chosen to ensure that the cascaded channel formed by the \gls{ris} provides a sufficiently strong link for redirecting the \glspl{ue}' signals to the \gls{bs}. 
Such placement can represent a maneuverable flying \gls{ris} mounted on an \gls{uav}, capable of dynamically establishing \gls{los} to \glspl{ue} that would otherwise encounter \gls{nlos} relative to a fixed-position \gls{bs}.

Lastly, both the \gls{bs} and the \glspl{ue} are equipped with single-antenna omnidirectional \glspl{tx}/\glspl{rx}, with their center frequency set to $5.9$~GHz, where \glspl{ris} are considered for deployment\cite{ETSI_RIS001}. For the generation of the wireless links, we leverage the capabilities of the \gls{quadriga}'s~\cite{jaeckel2014quadriga} statistical ray-tracing approach, which is based on channel sounding and geometry-based channel modeling. We observe the channels every $0.1$~s and select the 3GPP\textunderscore38.901\textunderscore UMa scenario for the channel generator. A number of \glspl{mpc} (denoted as $|N|$) are produced for each generated link, with the channel generation procedure based on the following flow adopted from~\cite{tsampazi2024system}:
\noindent
\begin{enumerate}

\item Generate scenario-specific \glspl{mpc} (denoted as $h_{xy}^\prime$) for each link, leveraging \gls{quadriga}.

\item Combine the \glspl{mpc} coherently to compute the channel gain for every link, following Eq.~\eqref{eq:mpc-def}:
\begin{equation}\label{eq:mpc-def}
\vspace{-0.12cm}
    \mathrm{h}^{\prime\prime}_{xy}=\sum_{j=1}^{|N|}\left|\textit{h}_{xy}^\prime\right| \cdot e^{j \varphi_{xy}^\prime},
\vspace{-0.2cm}
\end{equation}
\noindent
The \gls{mpc} gain is denoted as $\textit{h}^\prime_{xy}$, where $\text{\(x \in \{i, R\}, \ y \in \{R, A\}\)}$, and
$|\textit{h}^\prime_{xy}|$, $\varphi^\prime_{xy}$ are the respective  \gls{mpc} magnitude and phase.

\item Calculate the average of the channel gains (i.e., $\mathrm{h^{\prime\prime}_{xy}}$) over $100$ different channel realizations to determine the link's final channel gain (i.e., $h_{iA}$).

\item Multiply the computed channels by their corresponding steering vectors, denoted as $\mathbf{h}_{RA}^{\prime\prime\prime}$ and $\mathbf{h}_{iR}^{\prime\prime\prime}$, to obtain the respective final channel gains (i.e., $\mathbf{h}_{RA}$, $\mathbf{h}_{iR}$).

\item Calculate the overall channel power gain for the \gls{ue} $i$ to \gls{bs} link as follows $G_{i}=|h_{iA}+\mathbf{h}_{RA}^H \boldsymbol{\Theta} \mathbf{h}_{iR}|^2$.

\end{enumerate}

\section{Optimizing Wireless Propagation with RIS}\label{opt}

In this work, the wireless environment is not treated as an uncontrollable entity but is instead subject to optimization. For this purpose, we leverage the capabilities of \glspl{ris} to reshape the wireless propagation by effectively adapting the phase shifts of the \gls{ris} elements. For these reasons, we formulate the following optimization problem defined in Eqs.~(\ref{eq4a})-(\ref{eq4b}), as follows:
\begin{subequations}
\begin{equation}
\max_{\boldsymbol{\theta}} \sum_{i=1}^{|I|} |h_{iA}+\mathbf{h}_{RA}^H \boldsymbol{\Theta} \mathbf{h}_{iR}|^2
\label{eq4a}
\end{equation}
\begin{equation}
\text{s.t.\quad}0\leq\theta_m \leq 2\pi, \forall m\in M.
\vspace{-5pt}
\label{eq4b}
\end{equation}
\end{subequations}

It is noted that the solution to the aforementioned optimization problem, which takes place on the \gls{bs}'s side, is the effective phase shift vector of the \gls{ris} elements, denoted as $\boldsymbol{\theta^{*}}$. In order to maximize the overall channel power gain of the \glspl{ue}, the \gls{bs} only needs to calculate the effective phase shifts of the \gls{ris} elements (i.e., \( \boldsymbol{\theta^*} \)). Given the non-convexity of the aforementioned optimization problem, computing a globally optimal solution is challenging. To derive a closed-form phase-shift solution, we aim to achieve coherent signal alignment for the incoming signals from different transmission paths at the \gls{bs}. Indeed, the \gls{ue}'s channel power gain is maximized when the direct signal and the signal reflected by the \gls{ris} are perfectly aligned and coherently combined at the \gls{bs}'s \gls{rx}~\cite{li2020reconfigurable,diamanti2021energy}. This condition is generally satisfied when the phase shifts of the direct and cascaded signals are aligned as $\angle h_{iA} = \angle \left( \mathbf{h}_{RA}^{H} \mathbf{\Theta} \mathbf{h}_{iR} \right)$. Therefore, in the single-\gls{ue} case there's an \(1 \times |M|\) phase-shift vector \(\boldsymbol{\theta^*} = \angle \mathbf{v}\), where the optimal phase shifts of the \gls{ris} elements are given as follows: 
\begin{equation} \label{eq8}
\theta_{m}^{*}=\angle{h_{iA}}+\omega_{m}+\frac{2 \pi}{\lambda} d(m-1) \phi_{RA}, \forall m \in \ M.
\end{equation}

In the multi-\gls{ue} case, the solution to the maximization problem is also a linear combination of the phase shifts of the \gls{ris} elements, with the reflection-coefficient vectors \( \mathbf{v}_i \) being distinct for each \gls{ue}. Therefore, there exists a distinct reflection-coefficient vector \( \mathbf{v}_i = [v_{i,1}, \ldots, v_{i,|M|}] \in \mathbb{C}^{|M| \times 1} \) for each \gls{ue} \( i \) that maximizes its channel power gain. To determine the set of suitable weights for each \gls{ue} that will maximize the aggregate channel power gain across all \glspl{ue} as defined in Eq.~\eqref{eq4a}, we adopt the research methodologies proposed in~\cite{8970580,diamanti2021prospect}. For each \gls{ue} $i$, we define an appropriate weight factor \( w_i \in [0,1] \) and compute the linear combination of the overall
\glspl{ue}’ reflection-coefficients \( \mathbf{v}_i \) as follows:
\begin{equation}
\vspace{-0.18cm}
    \mathbf{v} = \sum_{i=1}^{|I|} w_i \mathbf{v}_i,
\label{ris-weigh-factor}
\vspace{-0.28cm}
\end{equation}
\noindent
such as the weight factors satisfy the condition \( \sum\limits_{i=1}^{|I|} w_i = 1\).
\noindent
In the multi-\gls{ue} case, the goal is to determine the optimal values of the weight factors  (i.e., $w_i$) for the reflection-coefficient vector (i.e., $\mathbf{v}$) of Eq.~\ref{ris-weigh-factor}, which maximize the overall channel power gain of the \glspl{ue}, as defined in Eq.~\eqref{eq4a}. Therefore, in order to compute the effective \gls{ris} elements phase shifts, the optimization problem defined in Eqs.~\eqref{eq4a}-\eqref{eq4b} is formulated as follows in Eqs.~\eqref{eq4-weights}-\eqref{eq4-weights3}:
\begin{subequations}
\begin{equation}
\max_{\textbf{w}} \sum_{i=1}^{|I|} |h_{iA}+\mathbf{h}_{RA}^H \boldsymbol{\Theta} \mathbf{h}_{iR}|^2
\label{eq4-weights}
\end{equation}
\begin{equation}
\text{s.t.\quad}0\leq w_i \leq 1, \forall i\in I 
\label{eq4-weights2}
\end{equation}
\begin{equation}
\sum_{i=1}^{|I|} w_i = 1
\vspace{-5pt}
\label{eq4-weights3}
\end{equation}
\end{subequations}
\noindent
Through backward induction~\cite{diamanti2021prospect}, the \gls{ris} reflection matrix \( \boldsymbol{\Theta} = \mathrm{diag}(e^{j \angle \mathbf{v}}) \) is computed to optimize the phase alignment. 
This approach ensures that the final \gls{ris} configuration \( \boldsymbol{\Theta} \) maximizes both the channel gain of each \gls{ue} and the total channel power gain across all \glspl{ue}. Given that 
\(\mathbf{w} = [w_1, \ldots, w_i, \ldots, w_{|I|}]\) is the vector of the \glspl{ue}' assigned weight factors, the optimization problem described in Eqs.~\eqref{eq4-weights}-\eqref{eq4-weights3} consists of a non-negative linear objective function and a set of constraints, which can be intuitively optimized to derive the optimal weights \( \mathbf{w}^* \). The derivation of the optimal weights \( \mathbf{w}^* \) leads to an effective \gls{ris} elements' reflection-coefficient vector 
\( \mathbf{v}^* = [v_1, \ldots, v_m, \ldots, v_{|M|}] \), which subsequently determines the corresponding phase shifts of the \gls{ris} elements, expressed as
\( \boldsymbol{\theta}^* = [\theta_1, \ldots, \theta_m, \ldots, \theta_{|M|}] \).

\section{Resource Allocation Strategies}\label{strategies}

We consider three different case studies, encompassing both single and multi-\gls{ue} scenarios, which are evaluated under various resource management policies in terms of \gls{prb} allocation and scheduling profile selection. 

\textbf{Case Study A}. Following the requirements outlined in~\cite{ETSI_RIS001}, which specify that existing \gls{ris} panels can host a minimum of $10$ elements, we evaluate three network configurations, where the number of \gls{ris} elements increases by orders of magnitude. 
We consider a single \gls{ue} allocated to the \gls{embb} network slice, which is assigned $\sim1/3$ of the available bandwidth resources. The \gls{csi} is varied to evaluate the contribution of \gls{ris} to network performance given the same resource availability, which is considered fixed. The results of this analysis are provided in Section~\ref{singleue-seca}. 

\textbf{Case Study B}. Considering the trade-off between hosting thousands of \gls{ris} elements and maintaining operational simplicity, we focus on a \gls{ris} with $100$ elements, serving multiple \glspl{ue}, which are distributed across the \gls{embb} and \gls{urllc} network slices. By dividing the available spectrum between the two, 
 we aim to demonstrate how a \gls{ris} can significantly enhance network performance, improve \gls{ue} satisfaction, and increase \gls{prb} utilization based on the capacity demands of each slice. The findings of this investigation are provided in Section~\ref{multiue-secb}.

\textbf{Case Study C}. This case study focuses on an \oran infrastructure where an xApp controls the scheduling profile selection of the \gls{bs}'s \gls{du}. The xApp updates the scheduling profile of the \gls{bs}'s scheduler through a control loop by alternating its policy given the following options: \gls{rr}, \gls{wf}, and \gls{pf}. A multi-\gls{ue} use case is considered, where \gls{embb} \glspl{ue} are provisioned with $90\%$ of the available bandwidth, while \gls{urllc} \glspl{ue} are allocated the remaining $10\%$ of the spectrum resources. With this case study, we aim to study the impact of an xApp frequently changing the scheduling profile from fair scheduling (i.e., \gls{rr}) to spectral efficiency-driven (i.e., \gls{wf}), or balanced scheduling (i.e., \gls{pf}), analyzing its effect on resource allocation and overall network performance in both \gls{ris}-assisted and non-\gls{ris}-assisted topologies. The results of this exploration are provided in Section~\ref{multiue-secc}.

\section{Experimental Setup}\label{expsetup}

To experimentally evaluate the \gls{ris}-assisted channels within an \oran ecosystem, we leverage the capabilities of OpenRAN Gym, an experimental toolbox for the end-to-end development, implementation, and testing of diverse solutions---including, but not limited to, \gls{ai}/\gls{ml} applications---within \oran. The software offers capabilities such as \gls{ran} and core network deployments using the srsRAN~\cite{gomez2016srslte} protocol stack. It supports large-scale data collection, testing, and fine-tuning of \gls{ran} functionalities by integrating open \glspl{api} 
for slicing and scheduling control, as well as \glspl{kpm} collection. Additionally, the software incorporates an \oran-compliant control architecture for executing xApps in the near-real-time \gls{ric}. Through the E2 interface, which bridges the \gls{ran} and \gls{ric}, along with its \glspl{sm}~\cite{polese2023understanding}, the streaming of \glspl{kpm} from the \gls{ran} and the execution of control actions by the xApps are enabled.

We deploy OpenRAN Gym on Colosseum, which features $128$~\glspl{srn} consisting of pairs of Dell PowerEdge R730 servers and NI \gls{usrp} X310 \glspl{sdr}, supporting large-scale experimentation in diverse network deployments. The testbed’s \gls{mchem} component emulates the wireless environment by leveraging \gls{fpga}-based \gls{fir} filters. These filters simulate \gls{rf} conditions, including path loss, fading and attenuation based on models created through various frameworks (e.g., ray-tracing software, analytical models, or real-world measurements). Similarly, the Colosseum \gls{mgen} TCP/UDP traffic generator~\cite{mgen} emulates a variety of network traffic profiles (e.g., multimedia content) and demand distributions (e.g., Poisson, periodic).

\begin{figure}[b!]
  \vspace{-1em}
  \centering
  \includegraphics[width=\columnwidth]{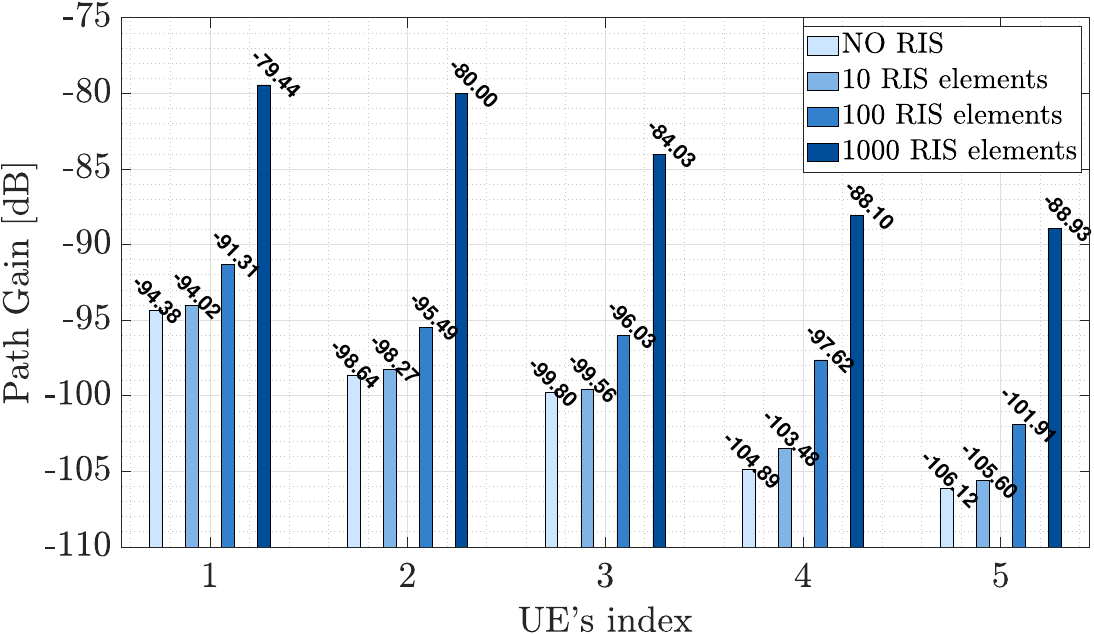} 
  \setlength\abovecaptionskip{-.1cm}
  \vspace{-0.3cm}
  \caption{Path Gains under varying numbers of \gls{ris} elements.}
  \label{fig:pathgainsris}
\end{figure}

We consider a cellular network comprising one \gls{bs} and up to $5$~\glspl{ue} distributed across $2$ network slices. These are: (i)~\gls{embb}, for high-traffic scenarios involving multimedia content and streaming applications; and (ii)~\gls{urllc}, designed for time-critical applications such as vehicle coordination in \gls{cv2x} environments. The maximum bandwidth considered in this work is is set to $10$\:MHz (i.e., $50$ \glspl{prb}) and is divided among the $2$ slices. Traffic is slice-based and generated according to the following specifications: \gls{embb} \glspl{ue} request a constant bitrate of $4$\:Mbps, while \gls{urllc} \glspl{ue} generate Poisson traffic at $89.3$\:kbps. In terms of physical deployment, the \gls{bs} is placed at the point $(x_A=25,y_A=50,z_A=25)$ [m] of the three-dimensional space. 
Their distance from the \gls{ris} is given as $d_{i,R}=[20, 27, 37, 58, 66]$~[m] and all are uniformly distributed around the \gls{ris}'s reference point, denoted as $(x_R=30,y_R=40,z_R=20)$ [m]. 

We install the generated links in Colosseum. The Colosseum scenario includes the path gain (i.e., a complex \gls{fir} coefficient) and the \gls{toa} values, 
which are are calculated based on the Euclidean distance between the active entities (i.e., \gls{bs}, \gls{ris}, and \glspl{ue}), using the speed of light. 
These values are used to generate the time-variant \gls{cir} for each node pair in the network.

\section{Experimental Evaluation\texorpdfstring{\protect\footnote{It is noted that all the results presented in Section~\ref{expevaluation} have been averaged over multiple repetitions of the experiments.}}{}}\label{expevaluation}

\begin{table}[b]
\vspace{-.55cm}
\centering
\renewcommand{\arraystretch}{1.05} 
\Huge
\caption{Catalog of Network Configurations}
\begin{adjustbox}{width=1\linewidth}
\begin{tabular}{@{}ccccccc@{}}
\toprule
\multicolumn{1}{c}{\shortstack{\textbf{\centering Config.} \\ \textbf{ID}}} & 
\multicolumn{1}{c}{ \shortstack{\textbf{Number} \\ \textbf{of \glspl{ue}}}} &
\multicolumn{1}{c}{ \shortstack{\textbf{\gls{ue}} \\ \textbf{ID(s)}}} &
\multicolumn{1}{c}{ \shortstack{\textbf{Slice} \\ \textbf{Type}}} &
\multicolumn{1}{c}{ \shortstack{\textbf{Bandwidth} \\ \textbf{[MHz]}}} &
\multicolumn{1}{c}{\shortstack{\textbf{\gls{ris}} \\ \textbf{Assisted}}} &
\multicolumn{1}{c}{\shortstack{\textbf{xApp} \\ \textbf{Enabled}}} \\
\midrule
\centering \textbf{I} & $1$ & $5$ & \text{\gls{embb}} & \text{$3.6$} & \text{No \gls{ris}} & \text{No}\\
\centering \textbf{II} & $1$ & $5$ & \text{\gls{embb}} & \text{$3.6$} & \text{$100$ \gls{ris} el.} & \text{No} \\
\centering \textbf{III} & $1$ & $1$ & \text{\gls{embb}} & \text{$3.6$} & \text{$10$ \gls{ris} el.}         & \text{No} \\
\centering \textbf{IV} & $1$ & $1$ & \text{\gls{embb}} & \text{$3.6$} & \text{$1000$ \gls{ris} el.}             & \text{No} \\
\centering \textbf{V} & $5$ & \text{$\{1, 2\}$ \& $\{3, 4, 5\}$} & \text{\gls{embb} \& \gls{urllc}} & \text{$5$ \& $5$} & \text{No \gls{ris}}         & \text{No}\\
\centering \textbf{VI} & $5$ & \text{$\{1, 2\}$ \& $\{3, 4, 5\}$}& \text{\gls{embb} \& \gls{urllc}} & \text{$5$ \& $5$} & \text{$100$ \gls{ris} el.}             & \text{No} \\
\centering \textbf{VII} & $5$ & \text{$\{1, 2\}$ \& $\{3, 4, 5\}$} & \text{\gls{embb} \& \gls{urllc}} & \text{$9$ \& $1$} & \text{No \gls{ris}} & \text{Yes}\\
\centering \textbf{VIII} & $5$ & \text{$\{1, 2\}$ \& $\{3, 4, 5\}$} & \text{\gls{embb} \& \gls{urllc}} & \text{$9$ \& $1$} & \text{$100$ \gls{ris} el.}             & \text{Yes}\\
\bottomrule
\end{tabular}
\end{adjustbox}
\label{table:net-topologies}
\end{table}

\begin{figure*}[t!]
\centering
\subfigure[]{\includegraphics[height=2.85cm]{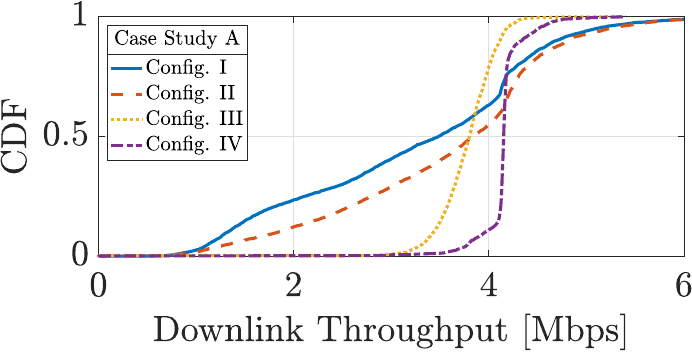}
\label{fig:singleue_thr}}
\hfil
\centering
\subfigure[]{\includegraphics[height=2.85cm]{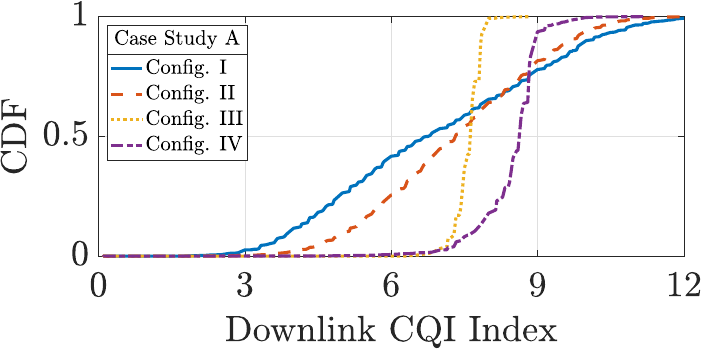}
\label{fig:singleuecqi}}
\hfil
\centering
\subfigure[]{\includegraphics[height=2.85cm]{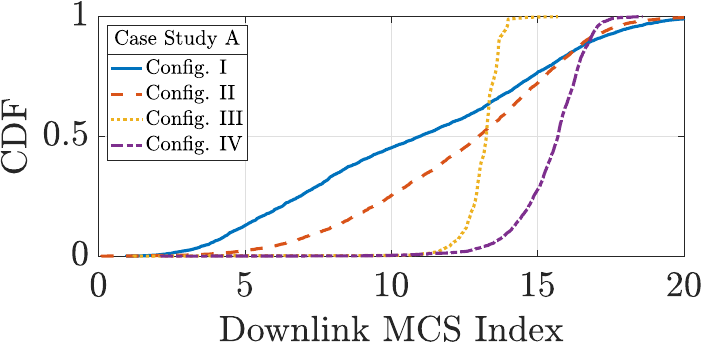}
\label{fig:singleuemcs}}
\setlength\abovecaptionskip{-.02cm}
\caption{Performance evaluation results for the single-\gls{ue} Case Study.}
\label{singleue-total}
\vspace{-1em}
\end{figure*}

\begin{figure*}[t!]
\vspace{-.5em}
\centering
\subfigure[]{\includegraphics[height=2.85cm]{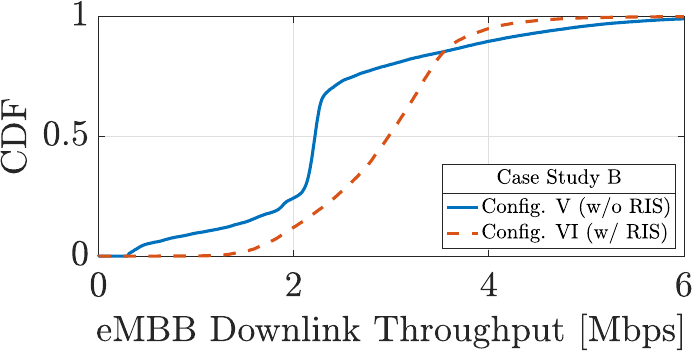}
\label{fig:multiue_thr}}
\hfil
\centering
\subfigure[]{\includegraphics[height=2.85cm]{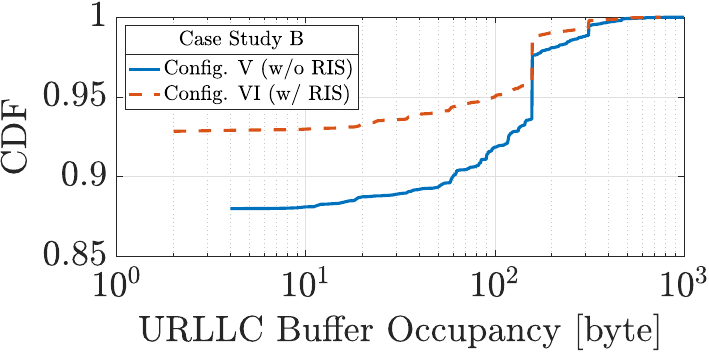}
\label{fig:multiuebuf}}
\hfil
\centering
\subfigure[]{\includegraphics[height=2.85cm]{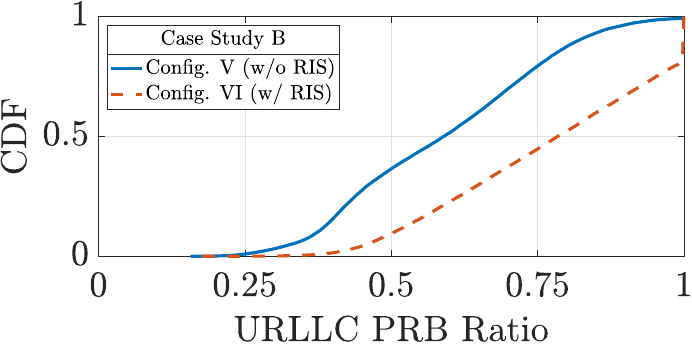}
\label{fig:multiueratio}}
\setlength\abovecaptionskip{-.02cm}
\caption{Performance evaluation results for the multi-\gls{ue} Case Study.}
\label{multiue-1}
\vspace{-1.5em}
\end{figure*}

\begin{figure*}[t!]
\centering
%
\subfigure[]{\includegraphics[height=3cm, width=4.4cm]{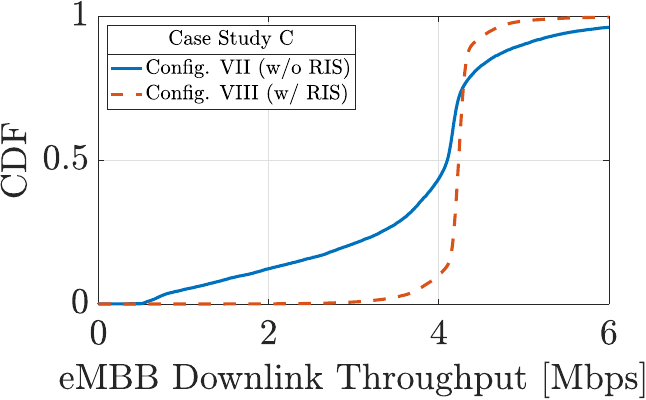}
\label{fig:multiue_thr_c}}
%
%
\hspace{-0.42cm}
\subfigure[]{\includegraphics[height=3cm, width=4.4cm]{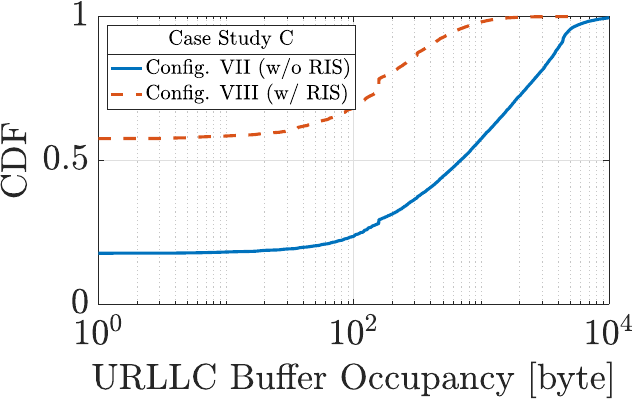}
\label{fig:multiuebuf_c}}
%
%
\hspace{-0.34cm}
\subfigure[]{\includegraphics[height=3cm, width=4.4cm]{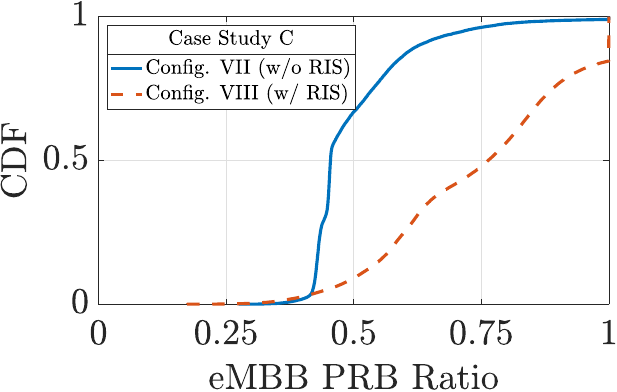}
\label{fig:multiueprb_c}}
%
%
\hspace{-0.22cm}
\subfigure[]{\includegraphics[height=3cm, width=4.4cm]{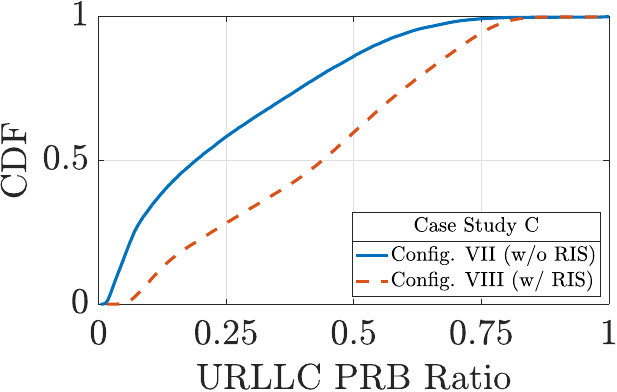}
\label{fig:multiue_case_c}}
\hspace{0.1\textwidth}
\setlength\abovecaptionskip{-.02cm}
\caption{Performance evaluation results for the multi-\gls{ue} \oran  Case Study.}
\label{multiue-2}
\vspace{-1.5em}
\end{figure*}

\begin{figure}[t!]
\centering
%
\subfigure[]{\includegraphics[height=2.8cm,width=4.1cm]{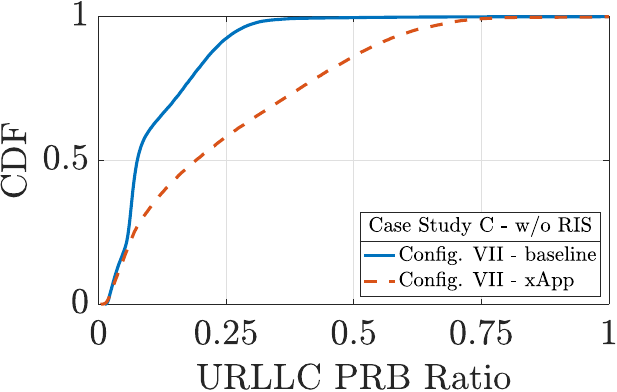}
\label{fig:baseline1}}
%
\hspace{-0.25cm}
\centering
\subfigure[]{\includegraphics[height=2.8cm,width=4.1cm]{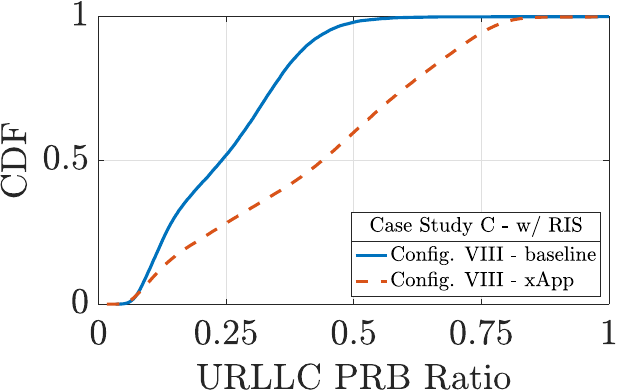}
\label{fig:baseline2}}
\setlength\abovecaptionskip{-.02cm}
\caption{Performance evaluation results for the multi-\gls{ue} Case Study with and without the inclusion of the \texttt{Sched} xApp in \oran.}
\label{baseline-total}
\vspace{-1.5em}
\end{figure}

In Fig.~\ref{fig:pathgainsris}, we present the path gains for each \gls{ue}, $i$, with and without a \gls{ris}, generated through the channel modeling process and the optimization procedure as described in Sections~\ref{channelmod} 
and Section~\ref{opt}, respectively.
All the \glspl{ue} are sorted in descending order, such that the \gls{ue} with the lowest index corresponds to the \gls{ue} with the best channel, while the \gls{ue} with the highest index corresponds to the \gls{ue} with the worst. Based on the reported results, the performance with as few as~$10$ \gls{ris} elements is nearly identical to that observed without the \gls{ris}. Improvements in path gains are evident across all \glspl{ue} when examining scenarios with $100$ or $1000$ \gls{ris} elements.

In Table~\ref{table:net-topologies}, we present a catalog of the various \textit{network configurations} evaluated within the scope of this work. Configs. \textbf{I}-\textbf{IV} focus on a single \gls{ue}, whereas configs. \textbf{V}-\textbf{VIII} address the multi-\gls{ue} case study, where scheduling decisions influence intra-slice resource allocation. Regarding configs. \textbf{VII}-\textbf{VIII}, an xApp tasked with reconfiguring the scheduling profile selection is evaluated. The xApp updates the scheduling profile of the \gls{bs} at a granularity of $1$~s, assigning one of three scheduling profiles (i.e., \gls{rr}, \gls{wf}, and \gls{pf}) to the \glspl{ue} of the \gls{embb} and \gls{urllc} slices, 
cycling through all possible combinations.

\subsection{Single-UE Case Study A}\label{singleue-seca}
In Fig.~\ref{singleue-total}, 
we compare the \gls{ue} with the best channel conditions, as shown in Fig.~\ref{fig:pathgainsris} (i.e., \gls{ue} with ID $1$), to the \gls{ue} with the worst (i.e., \gls{ue} with ID $5$). 
In Fig.~\ref{fig:singleue_thr}, focusing on Configs. \textbf{I} and \textbf{II} (Table~\ref{table:net-topologies}), which correspond to the \gls{ue} with the worst channel, we observe that the median throughput reported with $100$ \gls{ris} elements is $3.834$\:Mbps, representing an increase of $\sim10\%$ compared to the $3.486$\:Mbps observed with no \gls{ris}. Similarly, when focusing on the \gls{ue} with the best channel (Configs. \textbf{III} and \textbf{IV}), we observe that the median throughput reported with $1000$ \gls{ris} elements is $4.154$~Mbps, representing an $\sim10\%$ increase compared to the $3.8$~Mbps achieved with only $10$ \gls{ris} elements. 
It is reminded that according to the experimental results presented in Fig.~\ref{fig:pathgainsris}, the performance achieved with $10$ \gls{ris} elements is nearly identical to that observed without a \gls{ris} in the topology. 
Based on the aforementioned experimental evaluation, we observe that increases in the order of $10^{2}$ in the size of the panels (e.g., from no \gls{ris} to $10^{2}$ \gls{ris}, or from $10^{1}$ \gls{ris} to $10^{3}$ \gls{ris}) can result in a $\sim10\%$ increase in the reported throughput under the same resource availability (e.g., $18$~\glspl{prb} or $3.6$~MHz bandwidth). 

In Figs.~\ref{fig:singleuecqi} and~\ref{fig:singleuemcs}, we present the corresponding \gls{cqi} and \gls{mcs} values as reported by the protocol stack. The results in Fig.~\ref{fig:singleuecqi} demonstrate the channel improvement introduced by the \gls{ris} in the topology, particularly for the \gls{ue} with the best channel conditions (Configs. \textbf{III}-\textbf{IV}), which reports a $\sim13\%$ 
improvement in the median \gls{cqi} value (from $8$ to $9$). Similarly, the \gls{mcs} (Fig.~\ref{fig:singleuemcs}) increases from a median value of $13$ to $16$, enabling higher data rates given the same \gls{prb} availability. Regarding the \gls{ue} with the worst channel (Configs. \textbf{I}-\textbf{II}), the \gls{mcs} shows an increase of $\sim18\%$, rising from $11$ to $13$, while the \gls{cqi} remains relatively stable around $7$, with a modest increase of $\sim8\%$ for the \gls{ris} topology.
 
\subsection{Multi-UE Case Study B}\label{multiue-secb}
We examine the multi-\gls{ue} case study, where \glspl{ue} are allocated to specific slices based on their traffic demands.  To express \gls{ue} satisfaction (which corresponds to high resource utilization), we define a new metric, i.e., the \gls{prb} Ratio. Its definition is given as $\text{\small \gls{prb} Ratio}=\frac{\text{Sum of Granted \glspl{prb}}}{\text{Sum of Requested \glspl{prb}}}, \text{where \gls{prb} Ratio} \in [0, 1]$, and it represents the amount of allocated \glspl{prb} to the slice according to the channel and traffic conditions. The analysis focuses on Configs. \textbf{V} and \textbf{VI} (Table~\ref{table:net-topologies}), where the \glspl{ue} with the best reported channel conditions (Fig.~\ref{fig:pathgainsris}) are grouped together on the \gls{embb} slice, while the remaining \glspl{ue} are assigned to the \gls{urllc}. 
The \gls{wf} algorithm is selected for the scheduling profile of the \gls{embb} (which has the highest throughput requirements), while \gls{rr} is used for the \gls{urllc} (which has the lowest). 
Fig.~\ref{multiue-1} presents the complete performance evaluation results for the aforementioned case.
Fig.~\ref{fig:multiue_thr} illustrates the median \gls{embb} throughput for the \gls{ris} and non-\gls{ris} assisted topologies. The median throughput with a \gls{ris} is $2.973$\:Mbps, $\sim34\%$ higher than the $2.219$\:Mbps achieved without a \gls{ris}, under the same resource availability. Fig.~\ref{fig:multiuebuf} shows that both topologies (with and without a \gls{ris}) result in a median buffer occupancy of $0$ in the \gls{urllc}, indicating significantly reduced latency levels.\footnote{It is noted that since srsRAN does not directly measure latency, buffer occupancy is used as a proxy~\cite{tsampazi2024pandora}.} 
Fig.~\ref{fig:multiueratio} shows that although both topologies indicate near-zero latency, the inclusion of a \gls{ris} (Config. \textbf{VI}) leads to higher utilization efficiency of the available \glspl{prb} at the \gls{bs}'s scheduler.

\subsection{Multi-UE \oran Case Study C}\label{multiue-secc}
In Fig.~\ref{multiue-2}, focusing on Configs. \textbf{VII} and \textbf{VIII}, as detailed in Table~\ref{table:net-topologies}, we present the results of the experimental evaluation where the \texttt{Sched} xApps controls the \gls{bs}'s scheduling policy.
Fig.~\ref{fig:multiue_thr_c} shows that the median \gls{embb} throughput with a \gls{ris} in the topology is $4.234$\:Mbps, compared to $4.1$\:Mbps without a \gls{ris}. These high throughput values are reasonable, as the \gls{embb} \glspl{ue} are allocated $45$~\glspl{prb}, representing $90\%$ of the available resources. Fig.~\ref{fig:multiuebuf_c}, demonstrates that in the presence of a \gls{ris}, the median \gls{urllc} buffer occupancy is $0$, whereas in its absence, the corresponding value is $688$, indicating high latency levels. Therefore, the results show that the presence of $100$ \gls{ris} elements significantly reduces latency levels, regardless of the \gls{bs}'s scheduling policy. On the contrary, non-\gls{ris}-assisted topologies in resource-limited environments (i.e., with only $\sim10\%$ of the available bandwidth) experience performance degradation, even when the \gls{bs} frequently updates its scheduler from fair (e.g., \gls{rr}) to more efficient approaches (e.g., \gls{wf} or \gls{pf}) to accommodate the \glspl{ue} based on their distinct \gls{qos} demands.
Finally, Figs.~\ref{fig:multiueprb_c} and~\ref{fig:multiue_case_c} demonstrate that \gls{ris}-enabled topologies achieve higher resource utilization for both the \gls{embb} and \gls{urllc} slices. Figs.~\ref{fig:baseline1} and~\ref{fig:baseline2} compare the previously discussed topologies to the case where no xApp reconfigures the \gls{bs}'s scheduling policy. Notably, the keyword \texttt{baseline}, as shown in Fig.~\ref{baseline-total}, refers to srsRAN's default scheduling policy, which is \gls{rr}. Fig.~\ref{fig:baseline1} shows that updating the \gls{bs}'s scheduling profile increases resource efficiency for \gls{urllc} in the absence of a \gls{ris} (Config. \textbf{VII}). However, in the presence of a \gls{ris} (Fig.~\ref{fig:baseline2}), resource utilization and hence \gls{ue} satisfaction improve significantly (Config. \textbf{VIII}). Therefore, both the presence of \gls{ris} in the topology and frequent updates of the \gls{bs}'s scheduling policy enhance the resource utilization efficiency.

\section{Conclusions and Future Work}\label{conculsion}
In this work, we focused on an Open \gls{ran} and investigated the impact of various network management policies on the performance of \gls{ris}-assisted topologies, leveraging the capabilities of Open RAN Gym on Colosseum. Our experimental results demonstrate that a \gls{ris} deployment can enhance network throughput and resource utilization under the same bandwidth availability. In addition, our experimentation in an emulated \oran system, where the \gls{bs}'s scheduling policy is controlled by an xApp, demonstrates that \gls{ris}-assisted topologies achieve superior network performance and higher resource efficiency compared to their non-\gls{ris}-assisted counterparts, regardless of the scheduling policy employed. Part of our current and future work focuses on designing \gls{ai}/\gls{ml} solutions to dynamically control the selection of scheduling policies in such Open \glspl{ran}, leveraging \gls{csi} and traffic conditions in the decision-making processes. Extensions to the \oran control plane for carrying \gls{ris} weights, determined by the scheduler for real-time network adaptation, are also left for future work.
\vspace{-0.1cm}
\bibliographystyle{IEEEtran}
\bibliography{IEEEabrv,ref}
\end{document}